\documentclass[12pt]{article}
\pdfoutput=1
\usepackage{amsmath,amssymb}
\usepackage{graphicx}
\begin{document}
\def\slashchar#1{\setbox0=\hbox{$#1$} 
\dimen0=\wd0 
\setbox1=\hbox{/} \dimen1=\wd1 
\ifdim\dimen0>\dimen1 
\rlap{\hbox to \dimen0{\hfil/\hfil}} 
#1 
\else 
\rlap{\hbox to \dimen1{\hfil$#1$\hfil}} 
/ 
\fi}

\def\a{\alpha}
\def\b{\beta}
\def\c{\varepsilon}
\def\d{\delta}
\def\e{\epsilon}
\def\f{\phi}
\def\g{\gamma}
\def\h{\theta}
\def\k{\kappa}
\def\l{\lambda}
\def\m{\mu}
\def\n{\nu}
\def\p{\psi}
\def\q{\partial}
\def\r{\rho}
\def\s{\sigma}
\def\t{\tau}
\def\u{\upsilon}
\def\v{\varphi}
\def\w{\omega}
\def\x{\xi}
\def\y{\eta}
\def\z{\zeta}
\def\D{\Delta}
\def\G{\Gamma}
\def\H{\Theta}
\def\L{\Lambda}
\def\F{\Phi}
\def\P{\Psi}
\def\S{\Sigma}

\def\o{\over}
\def\beq{\begin{eqnarray}}
\def\eeq{\end{eqnarray}}
\newcommand{\gsim}{ \mathop{}_{\textstyle \sim}^{\textstyle >} }
\newcommand{\lsim}{ \mathop{}_{\textstyle \sim}^{\textstyle <} }
\newcommand{\vev}[1]{ \left\langle {#1} \right\rangle }
\newcommand{\bra}[1]{ \langle {#1} | }
\newcommand{\ket}[1]{ | {#1} \rangle }
\newcommand{\EV}{ {\rm eV} }
\newcommand{\KEV}{ {\rm keV} }
\newcommand{\MEV}{ {\rm MeV} }
\newcommand{\GEV}{ {\rm GeV} }
\newcommand{\TEV}{ {\rm TeV} }
\def\diag{\mathop{\rm diag}\nolimits}
\def\Spin{\mathop{\rm Spin}}
\def\SO{\mathop{\rm SO}}
\def\O{\mathop{\rm O}}
\def\SU{\mathop{\rm SU}}
\def\U{\mathop{\rm U}}
\def\Sp{\mathop{\rm Sp}}
\def\SL{\mathop{\rm SL}}
\def\tr{\mathop{\rm tr}}

\def\IJMP{Int.~J.~Mod.~Phys. }
\def\MPL{Mod.~Phys.~Lett. }
\def\NP{Nucl.~Phys. }
\def\PL{Phys.~Lett. }
\def\PR{Phys.~Rev. }
\def\PRL{Phys.~Rev.~Lett. }
\def\PTP{Prog.~Theor.~Phys. }
\def\ZP{Z.~Phys. }


\baselineskip 0.7cm

\begin{titlepage}

\begin{flushright}
ICRR-590-2011-7\\
IPMU11-0116
\end{flushright}

\vskip 1.35cm
\begin{center}
{\large \bf
Relatively Heavy Higgs Boson \\ in More Generic Gauge Mediation
}
\vskip 1.2cm
Jason L. Evans${}^{1}$, Masahiro Ibe${}^{1,2}$ and  Tsutomu T. Yanagida${}^{1}$
\vskip 0.4cm

${}^{1}${\it Institute for the Physics and Mathematics of the Universe (IPMU),\\ University of Tokyo, Chiba, 277-8583, Japan}\\
${}^{2}${\it Institute for Cosmic Ray Research, University of Tokyo, Chiba 277-8582, Japan }

\vskip 1.5cm

\abstract{
We discuss gauge mediation models where the doublet messengers and Higgs doublets are allowed to mix through a 	``charged'' coupling. The charged coupling replaces messenger parity as a means of suppressing flavor changing neutral currents without introducing any unwanted $CP$ violation.
As a result of this mixing between the Higgs doublets and the messengers,
relatively large $A$-terms are generated at the messenger scale. These large $A$-terms produce a distinct weak scale mass spectrum.
Particularly, we show that the lightest Higgs boson mass is enhanced and can be as heavy as $125$\,GeV for a gluino mass as light as $2$\,TeV.
We also show that the stops are heavier than that predicted by conventional gauge mediation models.
It is also shown that these models have a peculiar slepton mass spectrum.
}

\end{center}
\end{titlepage}

\setcounter{page}{2}

\section{Introduction}
A generic feature of models with gauge mediated supersymmetry (SUSY) breaking\cite{Dine:1981za, Dine:1981gu, Dimopoulos:1982gm,Affleck:1984xz,
Dine:1993yw,Dine:1994vc,Dine:1995ag}
is flavor-blind soft masses at the messenger scale.
Additionally, the $A$-terms at the messenger scale%
\footnote{In this paper, we define $A$-terms as the trilinear scalar couplings
divided by the corresponding Yukawa coupling constants.}
are generically predicted to be loop suppressed
compared to the sfermions masses. An important consequence of small $A$-terms at the messenger scale
is the lightest Higgs boson of the minimal supersymmetric standard model (MSSM)
is at the edge of the presently allowed mass region,
i.e. $m_{h^0} \lesssim  120$\,GeV
for $m_{\rm gluino}\lesssim 3$\,TeV\,\cite{Baer:2009dn}.%
\footnote{
Here, we roughly assume the reach of the LHC experiments
for $\sqrt{s}=14$\,TeV with the integrated luminosity $100$\,fb$^{-1}$.
}

In this letter, we investigate more generic gauge mediation models for which the $A$-terms are generated at one-loop without introducing any new flavor violation.
As a result of these large $A$-terms,  we show that the lightest Higgs boson is relatively heavy for a given gluino mass as compared with conventional gauge mediation models.
 Furthermore, we also show that the stop masses are predicted to be heavier than expected for such a heavy Higgs boson mass.
It is also shown that the models have a peculiar slepton mass spectrum.

The organization of the paper is as follows.
In section\,\ref{sec:model}, we discuss our more generic model of gauge mediation and explain the origin of
the large stop $A$-terms.
In section\,\ref{sec:higgs}, we discuss the distinct features of the mass spectrum, including the effects of the enhanced $A$-terms.
The final section is devoted to our conclusions and discussions.

\section{More Generic Gauge Mediation}\label{sec:model}
\subsection{Flavor blind models of gauge mediation}
Before discussing our more generic setup, we briefly review how conventional gauge mediation produces flavor-blind soft SUSY
breaking parameters at the messenger scale.
In most models, the messengers $(\Phi, \bar{\Phi})$ are assumed
to be a fundamental and anti-fundamental of the minimal grand unified gauge group, $SU(5)$.
The messengers couple directly with supersymmetry breaking in the superpotential
\begin{eqnarray}
 W = g Z \bar{\Phi}\Phi \ ,
\end{eqnarray}
where $g$ is some coupling constant and
\begin{eqnarray}
\label{eq:Z1}
g\vev{ Z} = M + F\theta^2.
\end{eqnarray}
We consider $g\vev{Z}$ to be a spurion and do not consider its origin in what follows%
\footnote{
See, for example, Refs.\,\cite{Ibe:2010jb,Evans:2011pz} for a recent discussion of models
of gauge mediation which include the SUSY breaking sector.
}.
Integrating out the messengers produces flavor-blind soft SUSY breaking parameters.

An implicit, but crucial, assumption made in the above discussion is that
the messengers do not directly couple to the MSSM matter fields.
In order to rationalize the above assumptions,
quantum numbers are assigned to the messengers
to distinguish them from the MSSM matter fields.
Most importantly, these quantum numbers distinguish the messenger doublets from the Higgs doublets. Without this distinction, the messengers would couple directly to the MSSM matter fields%
\footnote{
For simplicity, we have suppress the flavor indices in Eqs.\,(\ref{eq:FL}) and (\ref{eq:PROTON}).
}
\begin{eqnarray}
\label{eq:FL}
W = \rho_1{\Phi}_{\bar L} Q_L \bar{U}_R +\rho_2\bar{ \Phi}_{\bar L} Q_L \bar{D}_R
+ \rho_3\bar{\Phi}_{\bar L} L_L \bar{E}_R\ ,
\end{eqnarray}
leading to flavor-violating soft scalar masses\footnote{Tree level flavor violation from these new interactions could also be problematic unless the messenger scale is above about $10^5$ TeV.}. Furthermore, operators inducing rapid proton decay such as,
\begin{eqnarray}
\label{eq:PROTON}
 W =  \lambda_1\Phi_{D}Q_LQ_L + \lambda_2\bar{\Phi}_{ {D}}Q_LL_L\ ,
 \end{eqnarray}
could not be forbidden. Here, we have split the messengers into $\Phi = (\Phi_{D}, \Phi_{\bar L})$
and $\bar\Phi = (\bar{\Phi}_{D}, \bar{\Phi}_{\bar L})$ in accordance with
the MSSM gauge charges.
Forbidding these two types of operators, via quantum numbers, is crucial for building a successful model.%
\footnote{We assume that the messenger triplets, $\Phi_{ D}$ ($\bar{\Phi}_{ D}$)
have the same quantum numbers as the messenger doublets $\Phi_{\bar L}$ ($\bar{\Phi}_{\bar L}$)
under the symmetries which forbid the unwanted terms.
} This reasoning seems to eliminate the possibility of the Higgs doublets mixing with
the doublet messengers, $\Phi_{\bar L}$ and $\bar{\Phi}_{\bar L}$.

In supersymmetric theories, however, the Higgs doublets can mix with
the doublet messengers while the tree level operators in Eqs.\,(\ref{eq:FL}) and (\ref{eq:PROTON}) are forbidden.
This is accomplished by assuming that there is a $U(1)$ symmetry which was spontaneously broken
at some high energy scale by a single positively charged spurion field, $\phi_+$.
If the Higgs doublet is negatively charged under this $U(1)$, a combination of $\phi_+$ and $H_u$ can mix with the doublet messengers in the superpotential.
More explicitly, we may generalize the messenger sector as follows
 \begin{eqnarray}
 \label{eq:MIX}
 W = g Z  \bar\Phi\Phi
  +\frac{ \vev{\phi_+^2}}{\Lambda^2} Z\bar{\Phi} H_u
 \ ,
\end{eqnarray}
where $\Lambda$ denotes some high energy cutoff scale such as the Planck scale.
The unwanted operators in Eqs.\,(\ref{eq:FL}) and (\ref{eq:PROTON})
are forbidden if the charge assignments are as in Tab.\,\ref{tab:charge}.
Therefore, with the help of the ``charged coupling constant" $\vev{\phi_+}$,
the messengers and the Higgs pair can have a Yukawa interaction
in the superpotential without introducing any other phenomenological problems. This additional interaction will eventually lead to  mixing between the Higgs and the doublet messengers.
We note here that negatively charged couplings are not allowed in the superpotential because of its holomorphy (i.e. the so called SUSY-zero mechanism).
It is for this reason that the messengers are not allowed to have direct Yukawa interactions with the MSSM matter fields.
\begin{table}[t]
\caption{\sl\small
The charge assignments for the broken $U(1)$ symmetry are presented here.
We have used $SU(5)$ GUT representations for the MSSM matter fields,
i.e. ${\mathbf 10} = (Q_L,\bar{U}_R,\bar{E}_R) $
and ${\mathbf 5}^* = (\bar{D}_R,L_L) $.
We also show the charge of the right-handed neutrinos $\bar{N}_R$, which is need for the see-saw mechanism\,\cite{seesaw}.
These charge assignments forbid the unwanted interactions in Eq.\,(\ref{eq:FL}) and (\ref{eq:PROTON}) while allowing those in Eq.\,(\ref{eq:MIX}) along with the MSSM Yukawa interactions, the $\mu$-term, and mass terms for the right-handed
neutrinos.
}
\begin{center}
\begin{tabular}{|c|c|c|c|c|c|c|c|c|c|}
\hline
& $\phi_+$ & $H_u$ & $H_d$ & ${\mathbf 10}$ & ${\mathbf 5}^*$& $\bar{N}_R $
& $\Phi $ & $\bar\Phi$ & Z
\\
\hline
$U(1)$& $+1$ & $-2$ & $-3$ & $+1$ & $+2$ & $0$& $0$& $0$& $0$
\\
\hline
\end{tabular}
\end{center}
\label{tab:charge}
\end{table}%

From the above arguments, we find four classes of
models consistent with flavor constraints and rapid proton decay constraints;
\begin{itemize}
\item No mixings between the messengers and the Higgs pair.
\item The messenger $\Phi_{\bar L}$ mixes with $H_u$ with the help of a ``charged" coupling constant.
\item The messenger $\bar{\Phi}_{\bar L}$ mixes with $H_d$ with the help of a ``charged" coupling constant.
\item The messengers $\Phi_{\bar L}$ and $\bar{\Phi}_{\bar L}$ mix with $H_u$ and $H_d$, respectively,
with the help of ``charged" coupling constants.%
\footnote{ A similar model to the fourth possibility has been considered
based on a framework of the extra dimension\,\cite{Chacko:2001km},
where the operators causing rapid proton decay
are suppressed by brane separation, while
they are suppressed by the SUSY-zero mechanism in the present model as explained in the text.
}
\end{itemize}
The first class of models corresponds to conventional gauge mediation.
The second class which we name Type-II gauge mediation is a new class of models.
Below we will discuss the mass spectrum of Type-II gauge mediation. We discuss only Type-II gauge mediation since we are most interested in the mass of the lightest Higgs boson. However, these other two classes of models will have their own unique spectrum.

\subsection{Soft SUSY breaking masses in Type-II gauge mediation}
Now we examine in detail Type-II gauge mediation models. In Type-II gauge mediation models,
only $H_u$ mixes with the messengers. The superpotential for Type-II gauge mediation at the messenger scale is
\begin{eqnarray}
\label{eq:superpotential1}
 W = gZ\bar{\Phi}\tilde{\Phi}
 + g'Z\bar{\Phi}_{\bar L}  \tilde{H}_u
  + \tilde \mu \tilde{H}_uH_d  + \tilde{y}_{U ij} \tilde{H}_u Q_{Li} \bar{U}_{Rj}\ ,
\end{eqnarray}
where $\tilde\mu$ ($\propto \vev{\phi_+^5}/\Lambda^4$) is a dimensionful parameter,  $\tilde y_{Uij}$
is the usual $3\times 3$ Yukawa coupling matrix,
and we have replaced $\vev{\phi_+^2}/\Lambda^2$ by $g'$.
We have also placed tildes on $H_u$ and $\Phi_{\bar L}$ for later purposes
and have neglected the parts of the MSSM superpotential which are not relevant for our discussion.
Because of holomorphy of the superpotential, as explained above, the dangerous terms like $\Phi_{\bar L} Q_L \bar{U}_R$
are forbidden because a negatively charged couplings constant is not allowed.

To elicit the important low-scale phenomenon of these models, we change the field basis by the rotation
\begin{eqnarray}
\left(
\begin{array}{cc}
\tilde \Phi_{\bar L}\\
\tilde H_u
\end{array}
\right)=
\frac{1}{\sqrt{g^2+g'^2}}\left(
\begin{array}{cc}
g  & -g'     \\
g'  & g
\end{array}
\right)
\left(
\begin{array}{cc}
\Phi_{\bar L}\\
H_u
\end{array}
\right)\ .
\end{eqnarray}
In this new basis, the above superpotential becomes
\begin{eqnarray}
\label{eq:superpotential2}
 W = \bar{g} Z\bar\Phi {\Phi}  + \mu H_uH_d   + \mu' \Phi_{\bar L} {H}_d + y_{Uij} H_u Q_{Li} \bar{U}_{Rj}
 + y_{Uij}' \Phi_{\bar L} Q_{Li} \bar{U}_{Rj}\ ,
\end{eqnarray}
where the parameters are defined as
\begin{eqnarray}
 \bar{g} &=& \sqrt{g^2 + g'^2}\ ,\quad
 \mu = \frac{g}{\sqrt{g^2 + g'^2}}\tilde\mu\ , \quad
  \mu' = \frac{g'}{\sqrt{g^2 + g'^2}}\tilde\mu\ , \quad\cr
&&  y_{Uij} = \frac{g}{\sqrt{g^2 + g'^2}}\tilde y_{Uij}\ , \quad
  y_{Uij}' = \frac{g'}{\sqrt{g^2 + g'^2}}\tilde y_{Uij} \ .
\end{eqnarray}
This new basis is much better for low scale physics because the only heavy states are clearly $\Phi,\bar\Phi$.%
\footnote{
In the following arguments, we slightly change the definitions of the spurion VEV from
Eq.\,(\ref{eq:Z1}) to
\begin{eqnarray}
 \bar{g} \langle Z \rangle= M + F\theta^2 \ .
\end{eqnarray}
}
In this basis, the mixing angle between
the Higgs doublet and the messengers is suppressed by $O(\m/M)$, as compared to $O(g'/g)$ in the original. Since we will consider $g'/g\sim 1$, this basis is more suited for physics below the messenger scale.

Here, we reaffirm that the new flavor dependent interactions in this basis,
\begin{eqnarray}
 W = y_{Uij}' \Phi_{\bar L} Q_{Li} \bar{U}_{Rj}\ ,
\end{eqnarray}
are not dangerous.  Since these new flavor dependent interactions are aligned with the MSSM
Yukawa coupling, $y_U$, diagonalizing the Higgs Yukawa couplings will simultaneously diagonalize these additional Yukawa couplings.%
\footnote{In this sense, the Type-II gauge mediation
is a natural realization of the so called ``minimal flavor violation" scenario (see for example Ref.\,\cite{D'Ambrosio:2002ex}).
}
In the MSSM, we are free to chose one of the Yukawa couplings to be diagonal without any loss of generality. In this basis, it is clear that no new significant source of flavor violation is present. In the following discusssion, we choose the basis where ${\tilde y}_U$ is diagonal and neglect everything except the top Yukawa coupling,
\begin{eqnarray}
\label{eq:top}
W = y_t H_u Q_{L3} \bar{T}_R +  y_t' \Phi_{\bar L} Q_{L3} \bar{T}_R\ .
\end{eqnarray}
Not only are these interaction not dangerous, but it is these new interactions that give type-II gauge mediation its unique spectrum.

\subsubsection*{Tree-level mediation effect}
The third term of the superpotential in Eq.\,(\ref{eq:superpotential2})
leads to a soft SUSY breaking squared mass for $H_d$
at the ``tree-level".
That is, by integrating out the messengers, the down-type Higgs $H_d$ gets a tree-level soft squared mass,
\begin{eqnarray}
\label{eq:tree}
   m_{\bar H}^2 = - \mu'^2 \frac{F^2}{M^4-F^2}\ .
\end{eqnarray}
Here, $\mu'$ is assumed to be of the same order of magnitude as the $\mu$-term,
for $g/g'=O(1)$.
This contribution can be important in low scale gauge mediation where $F\sim M^2$.
However, as we push up the messenger scale this contribution falls off quickly.
Fortunately, this tree-level mediation does not play an important role
in most of the parameter space we are interested in.

\subsubsection*{The one-loop contribution to  $m_{Q}^2$ and $m_{T}^2$}
\begin{figure}[t]
\begin{center}
  \includegraphics[width=.7\linewidth]{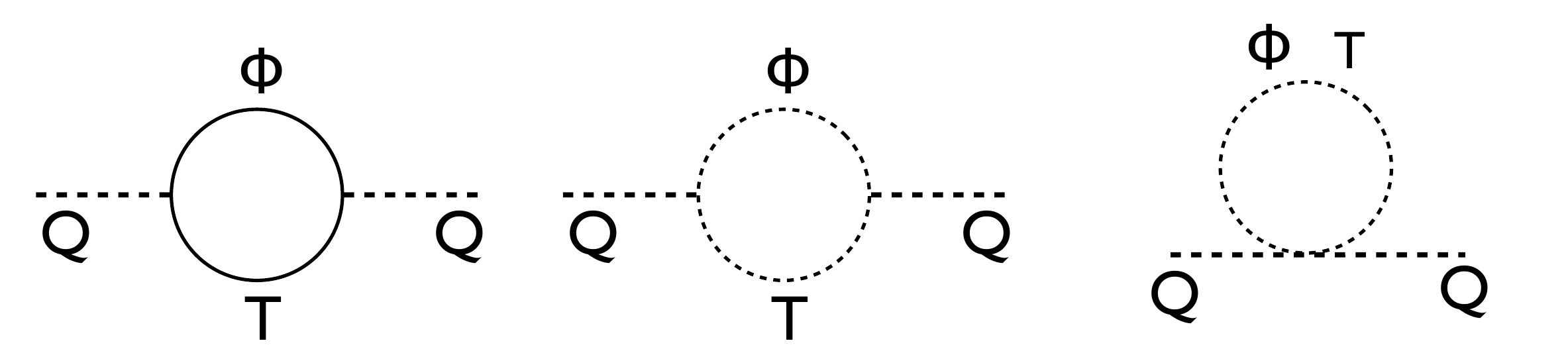}
\caption{\sl \small
The diagrams which are relevant for the soft squared mass
of $Q_L$.
The soft squared mass of $\bar{T}_R$ is obtained
by exchanging the $Q$'s and $T$'s in the diagrams.
}
\label{fig:scalar}
\end{center}
\end{figure}
Now, let us discuss the one-loop soft squared mass of $Q_{L3}$
due to the last interaction term in Eq.\,(\ref{eq:top}).
From the diagrams in Fig.\,\ref{fig:scalar}, we obtain
a one-loop soft squared mass for $Q_{L3}$;
\begin{eqnarray}
\delta m_{Q_3}^2
&=&  \frac{y_t'^2}{32\pi^2}\frac{F^2}{M^2}
\left(
\frac{
(2+x)\log(1+x)
+
(2-x)\log(1-x)
}{x^2}
\right)\ ,
\end{eqnarray}
where we have defined,
\begin{eqnarray}
x = \frac{F}{M^2}\ .
\end{eqnarray}
In a similar way, we obtain the soft mass of $\bar{T}_R$;
\begin{eqnarray}
\delta m_{\bar{T}}^2 =  2 \times \delta m_{Q_3}^2 \ .
\end{eqnarray}

It should be noted that these one-loop contributions to the stop squared masses are negative\,\cite{Evans:2010kd}.
Thus, one might worry that these one-loop negative contributions
dominate the positive but two-loop gauge mediated contributions since $y_t' \simeq 1$.
This is, however, not the case for $x\ll 1$ since these one-loop contributions are suppressed
by additional factors of $x$ ,
\begin{eqnarray}
\label{eq:Oneloop}
\delta m_{\bar T}^2
\simeq -  \frac{y_t'^2}{48\pi^2}\frac{F^2}{M^2} \frac{F^2}{M^4}\ , \quad (x\ll1 )\ .
\end{eqnarray}
The two-loop dominance can be seen explicitly by comparing the above contribution to the stop mass with the traditional gauge mediated contribution,%
\footnote{
We neglected the gauge mediated contributions other than the ones from the strong interactions.
}
\begin{eqnarray}\label{eq:GMSB}
 m_{Q,T}^{2} \simeq \frac{8}{3}\left(\frac{\alpha_3}{4\pi}\right)^2 \frac{F^2}{M^2}\ , \quad (x\ll1 ).
\end{eqnarray}
The one-loop contribution is subdominant and does not lead to a tachyonic stops mass as long as,
\begin{eqnarray}
\frac{F}{M^2} \ll 2 \sqrt{2}\times \frac{\a_3}{y_t'} \ .
\end{eqnarray}
This condition can be easily satisfied, even for $y_t' \simeq 1$, as long as the messenger scale is not too low.

\begin{figure}[t]
\begin{center}
  \includegraphics[width=.55\linewidth]{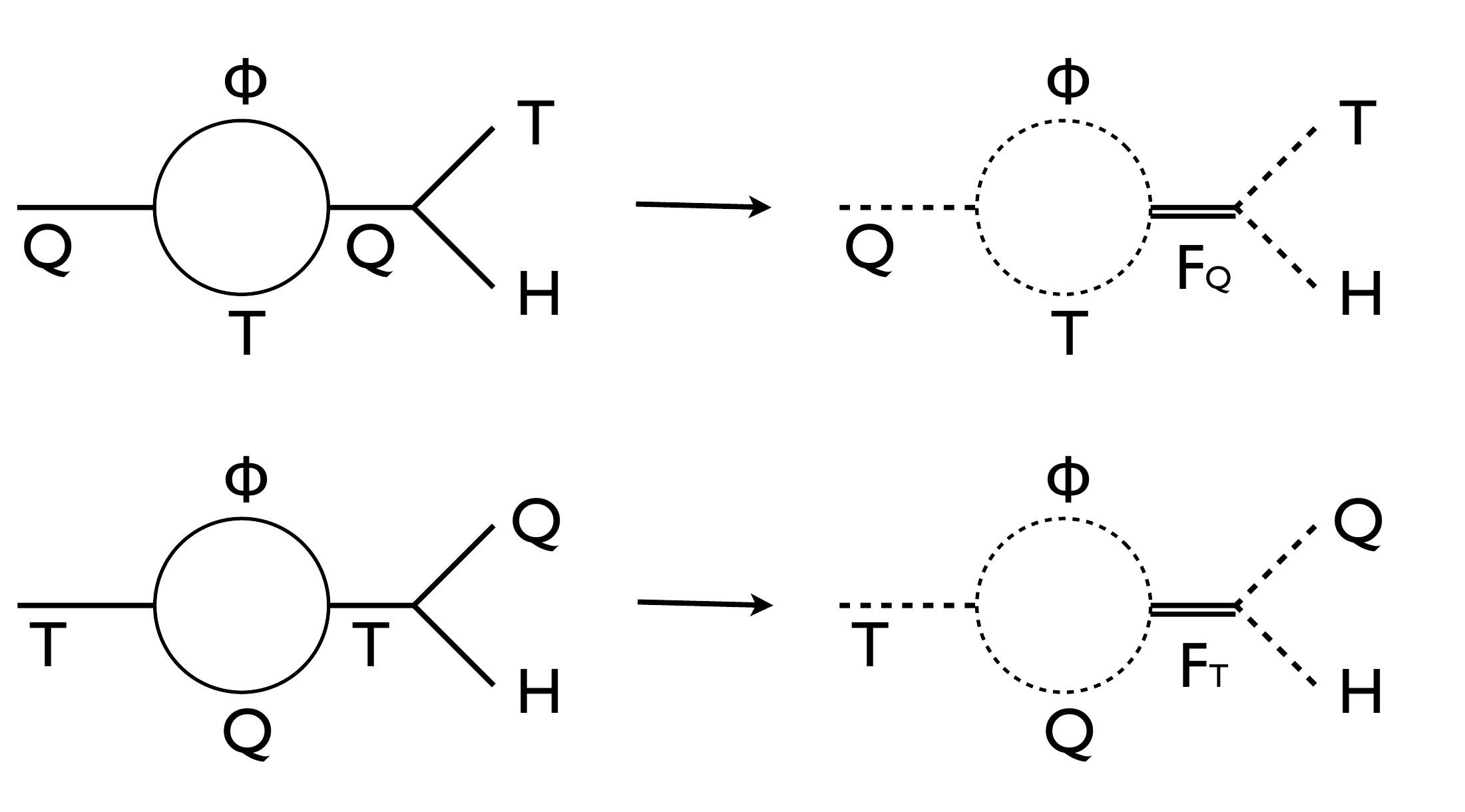}
\caption{\sl \small
The diagrams which are relevant for the $A$-terms.
In terms of supergraphs, the $A$-term is generated
as a result of the wave function renormalization,
which reduces to the 1PI diagrams in component
graphs.
}
\label{fig:Aterm}
\end{center}
\end{figure}

\subsubsection*{The one-loop contributions to $A$-terms}
In Type-II models, $A$-terms are also generated at the one-loop level (see Fig.\,\ref{fig:Aterm}).
The resultant $A$-term for the stops is given by
\begin{eqnarray}
 A_t  = -\frac{3}{32\pi^2} y_t'^2 \frac{F}{M}\frac{1}{x}\log\left(\frac{1+x}{1-x}\right)\ .
\end{eqnarray}
The $A$-term for the sbottoms is also given by, 
\begin{eqnarray}
 A_b  = -\frac{1}{32\pi^2} y_t'^2 \frac{F}{M}\frac{1}{x}\log\left(\frac{1+x}{1-x}\right)\ .
\end{eqnarray}
In contrast to the one-loop soft squared masses,
the $A$-terms have no $x$ suppression in the limit $x\ll1$,
\begin{eqnarray}
\label{eq:Aterm2}
 A_t \simeq - \frac{3 y_t'^2} {16\pi^2} \frac{F}{M}\ , \quad (x\ll1 ).
\end{eqnarray}
Thus, the one-loop contribution to the $A$-terms can be sizable even when
the messenger scale is very high (i.e. $x\ll1$) if $y_t' \simeq 1$.
As we will see shortly, these relatively large $A$-terms push up the mass of the lightest Higgs boson
significantly.

\subsubsection*{The two-loop contribution to $m_Q^2$ and $m_T^2$}
Finally, let us discuss the two-loop contributions to $m_Q^2$,  $m_T^2$ and $m_{H_u}^2$
from the last interaction term in Eq.\,(\ref{eq:top}).
Unlike the one-loop contributions to $m_Q^2$ and $m_T^2$,
the two-loop contributions
are not suppressed in the limit of $x\ll 1$.%
\footnote{
We appreciate S.~Shirai for pointing out the unsuppressed two-loop contributions.
}
The leading two-loop contributions can easily be extracted by analytically continuing the wave function renormalization factor into superspace\,\cite{Giudice:1997ni}
which leads to\footnote{ Our results for the two loop contribution disagrees
with the results given in Ref.\,\cite{Chacko:2001km}.}
\begin{eqnarray}
\label{eq:Twoloop}
 \delta m_{Q_3}^2 &=& \frac{y_t'^2}{128\pi^4}\left(3y_t'^2 +3y_t^2
 - \frac{8}{3}g_3^2
 -\frac{3}{2} g_2^2
 -\frac{13}{30} g_1^2\right)\frac{F^2}{M^2}\ , \cr
 \delta m_{\bar T}^2 &=& \frac{y_t'^2}{128\pi^4}\left(6y_t'^2  + 6y_t^2
 + y_b^2
 - \frac{16}{3}g_3^2
 -3 g_2^2
 -\frac{13}{15} g_1^2\right)\frac{F^2}{M^2}\ ,\cr
  \delta m_{\bar B}^2 &=& - \frac{y_b^2y_t'^2 }{128\pi^4}\frac{F^2}{M^2}\ , \nonumber \\
  \delta m_{H_u}^2 &=& - 9\frac{y_t^2y_t'^2 }{256\pi^4}\frac{F^2}{M^2}\ , \nonumber \\
    \delta m_{H_d}^2 &=& - 3\frac{y_b^2y_t'^2 }{256\pi^4}\frac{F^2}{M^2}\ ,
\end{eqnarray}
where $y_b$ is the bottom Yukawa coupling constant and $m_{\bar B}^2$
is the soft squared mass of the right-handed sbottom.
It should be noted that the two-loop contributions to $m_{H_u}^2$ is negative,
and can dominate the gauge mediated contributions if $y_{t}' \simeq 1$.
Since the Higgs doublets can have a large supersymmetric mass, $\mu$, the negative value of $m_{H_u}^2$ does not
lead to a vacuum stability problem.

Several comments are in order.
First, let us point out that there are no $CP$-phases in Eq.\,(\ref{eq:superpotential2}).
All the phases of $\tilde y_t$ and $\mu$, except the CKM phase, can be eliminated by rotating the MSSM fields in the standard way.
The phases of the new couplings $g$ and $g'$ can also be eliminated
by appropriately rotating $\Phi$ and $\bar\Phi$.
Therefore, the SUSY $CP$-problem is also absent.
It should also be noted that Type-II models do not lead to a large $B\mu$-term,
since the diagrams which leads to the $B\mu$-term require two insertion of $\mu'$.
Thus, these models are also free from the so-called $B\mu$-term problem.%
\footnote{
Although we do not discuss the solution to the so-called $\mu$-problem in this paper,
we may further extend the Higgs sector so that $\mu$ and $B$ are
generated with similar size in a $CP$-safe way\,\cite{Evans:2010ru}.}

\section{The Spectrum of the Model}\label{sec:higgs}
In this section, we show the distinctive features of the spectrum
of the Type-II gauge mediation models.
\subsection{The heavy lightest Higgs region}
\begin{figure}[t]
\begin{center}
\begin{minipage}{.49\linewidth}
  \includegraphics[width=.8\linewidth]{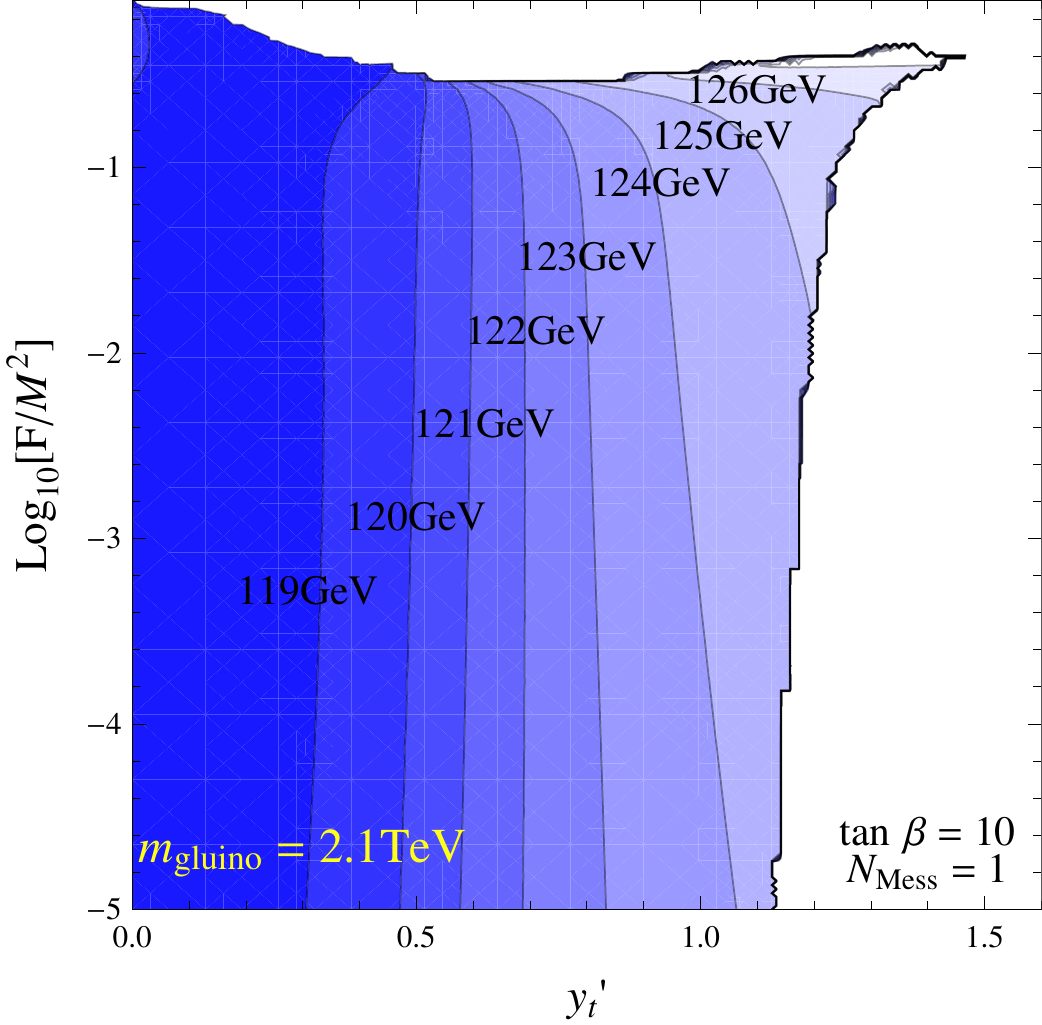}
  \end{minipage}
  \begin{minipage}{.49\linewidth}
  \includegraphics[width=.9\linewidth]{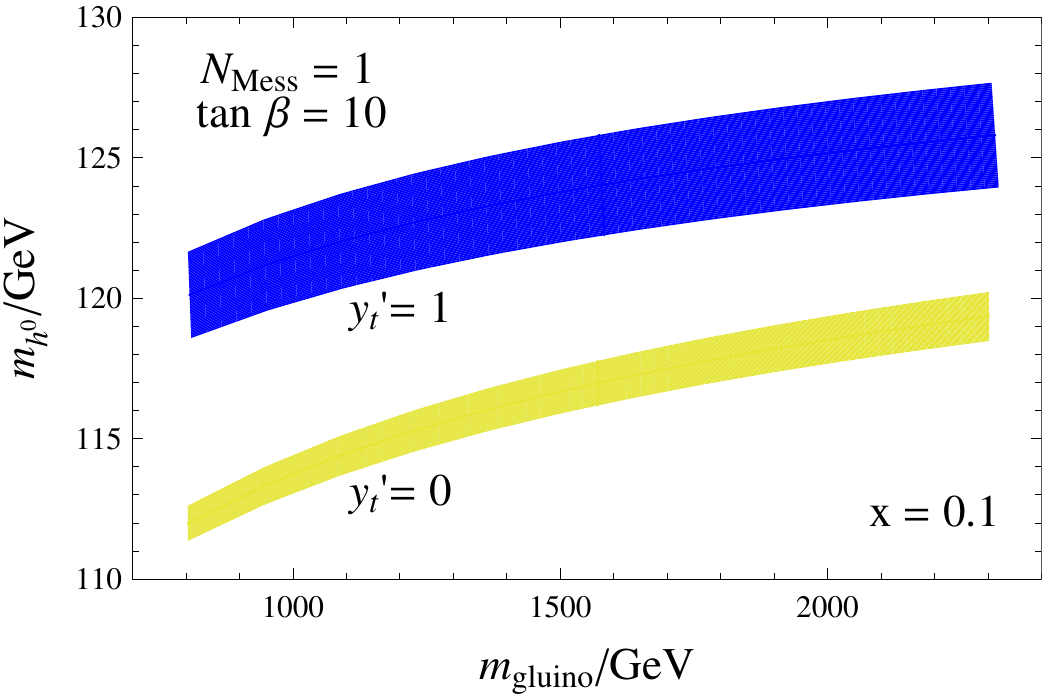}
  \end{minipage}
\caption{\sl \small
Left) The contour plot of the lightest Higgs boson mass
for $m_{\rm gluino}\simeq 2.1$\,TeV, $\tan\beta = 10$,
$N_{\rm mess} = 1$ and $m_{t} = 173.2$\,GeV.
The blank region for $x \simeq 1$ is mainly excluded because of tachyonic stop masses.
The blank region for $y_{t'}\gtrsim 1$ and $x \ll 1$ is excluded
by the tachyonic slepton masses.
Right)
The lightest Higgs boson mass for a given gluino mass.
The blue band corresponds to the parameters, i.e. $y_t' =1$  and $x = 0.1$.
We also show the upper bound for conventional models of gauge mediation, i.e. $y_t' = 0$ (yellow band).
The upper and the lower boundary of the each band corresponds to
the upper and the lower limits of the current world average top mass,
$m_t = 173.2\pm 0.9$\,GeV\,\cite{Tev}.
The relatively broad band in Type-II gauge mediation
represents the fact that our reference point $(y_t' =1.0, x=0.1)$
is not the optimal Higgs mass point for a given value of $m_{\rm top}$.
}
\label{fig:Higgs}
\end{center}
\end{figure}
As we have shown in the previous section,
an interesting feature of Type-II gauge mediation
is the relatively large stop $A$-terms which are generated at the one-loop level.
With a relatively large $A$-term, the lightest Higgs boson mass,
which receives important SUSY breaking corrections via the top-stop loop diagrams\,\cite{Higgs},
is pushed up to
\begin{eqnarray}
\label{eq:Higgs}
 m_{h^0}^2 \lesssim m_Z^2 \cos^22\beta + \frac{3}{4 \pi^2}y_t^2 m_t^2 \sin^2{\beta}
\left( \log \frac{m_{\tilde t}^2 }{m_{t}^2 }+
\frac{A_t^2}{m_{\tilde t}^2}
-\frac{A_t^4}{12m_{\tilde t}^4}
\right)\ .
\end{eqnarray}
Here, $m_{Z}$ and $m_t$ are the masses of the $Z$-boson and top quark,
respectively, and $\tan\beta$ is the ratio of the two vacuum expectation values of the Higgs doublets.%
\footnote{In the above expression, we have neglected the stop mixing from the $\mu$-term since
it is suppressed for $\tan\beta \gtrsim 10 $.}
The above expression for the Higgs mass is maximized for an $A$-term of order $A_t \simeq \sqrt{6} \times m_{\tilde t}.$
By comparing Eqs.\,(\ref{eq:GMSB}), Eq.\,(\ref{eq:Aterm2}), and Eq.\,(\ref{eq:Twoloop}),
we see that the lightest Higgs boson receives large $A$-term contributions
for $y_t' \simeq 1$.

In Fig.\,\ref{fig:Higgs}, we show a contour plot of the lightest Higgs boson mass
as a function of $y_t'$ and $F/M^2$ for a gluino mass of $2.1\,{\rm TeV}$, $\tan\beta=10$
and the number of messenger $N_{\rm mess}=1$ (left panel).
To calculate the weak scale soft masses and Higgs boson mass, we have used {\it SoftSusy}\,\cite{Allanach:2001kg}.%
\footnote{
The uncertainty of the lightest Higgs boson mass
is estimated to be about $2-5$\,GeV\,\cite{Allanach:2004rh}.
}
The figure shows that the Higgs mass becomes large for $y_{t}' \simeq 1$ as expected.

The blank regions in Fig. \ref{fig:Higgs} are excluded.  For $x\simeq 1$, the stop mass is tachyonic (or too light) and for $y_t' \gtrsim 1$ and $x\ll1$ the right handed slepton masses are tachyonic\footnote{The origin of the tachyonic slepton is discussed below.} (or too light).
Within the allowed region, we find that the vacuum stability condition\,\cite{Kusenko:1996jn},
\begin{eqnarray}
   A_t^2 + 3 \mu^2 < 7.5\, (m_{{\tilde t}_L}^2 + m_{{\tilde t}_R}^2 ) \ ,
\end{eqnarray}
is always satisfied and the relatively large $A$-terms do not cause vacuum instability problems
in Type-II models.

The right panel of Fig.\,\ref{fig:Higgs} shows the lightest Higgs boson mass
for $y_t' = 1$ and $x  = 0.1$.
From this figure, we see that the mass is raised by about $10$\,GeV
compared to the Higgs boson mass of conventional gauge mediation.
As a result of this enhancement, the Higgs boson can be heavier than $120$\,GeV for a relatively light gluino mass,
$m_{\rm gluino}\sim 1$\,TeV.
The figure also shows that the gluino is well within the reach of the LHC experiments
even for a relatively heavy Higgs boson mass, i.e. $m_{h^0}\simeq125$\,GeV.

\begin{figure}[t]
\begin{center}
  \includegraphics[width=.55\linewidth]{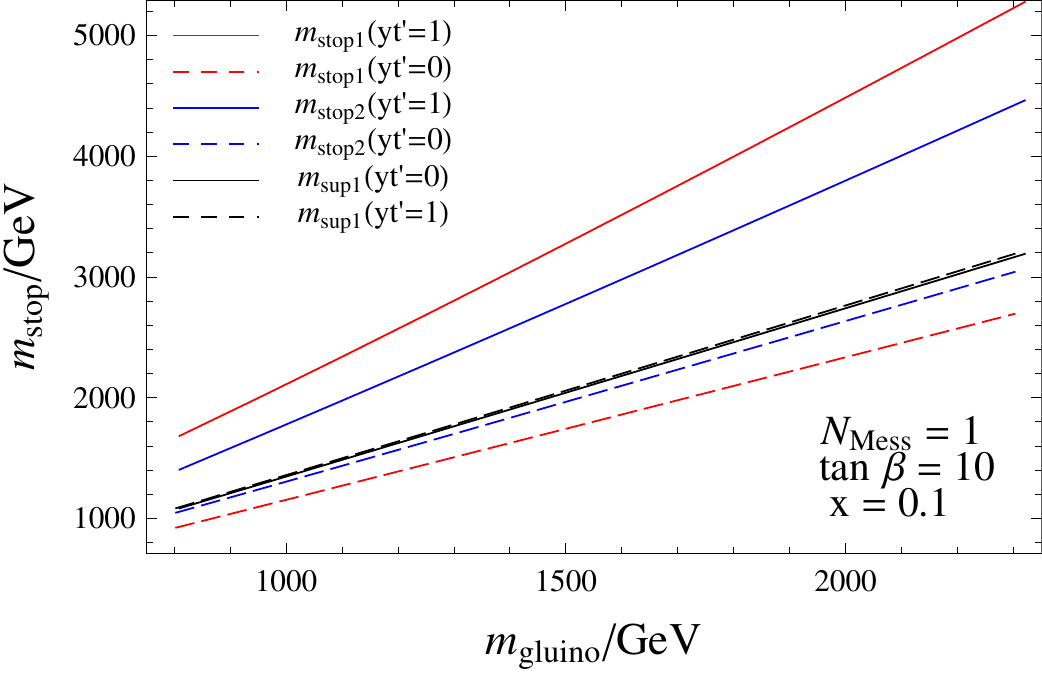}
\caption{\sl \small
The masses of the stops in Type-II models (solid lines) for $y_t' = 1$ and  $x = 0.1$
and for conventional models of gauge mediation (dashed lines).
The red and blue lines correspond to the masses of two stops.
For comparison, we show the sup masses for both models (black lines), however, there
is no difference in these masses for the two types of models.
In this figure, we have used $m_t = 173.2$\,GeV.
}
\label{fig:stop1}
\end{center}
\end{figure}

In addition to a relatively heavy lightest Higgs boson, we also expect that
the stops are heavier  than those of minimal gauge mediation.
This stop mass enhancement is mainly
due to the two-loop contribution to the stops in Eq.\,(\ref{eq:Twoloop}).
In Fig.\,\ref{fig:stop1}, we plot the stop masses for a representative point with a large Higgs mass\footnote{The Higgs mass can be further enhanced beyond what is shown. However, this may lead to a Landau pole below the Plank scale.},
$(y_t' =1, x = 0.1)$.
The figure shows that both stops are predicted
to be heavier than those in conventional gauge mediation models ($y_t'  = 0$).
These mass features provide important clues for probing Type-II gauge mediation
models at the LHC.%
\footnote{Detailed phenomenological analysis will be given elsewhere.}

The models also predict a peculiar slepton mass spectrum in the region where lightest Higgs mass is relatively heavy.
This peculiar slepton spectrum is caused by the renormalization group evolution of the sleptons,
\begin{eqnarray}
 \frac{d}{dt} m_{\rm slepton}^2 = -\sum_{a=1,2}8C_a\frac{g^2_a}{16\pi^2} |M_{a}|^2
 +\frac{1}{8\pi^2}\frac{3}{5} Y g_1^2 {\cal S}\ ,
\end{eqnarray}
where $M_a$ denote the gaugino masses,
$C_{2} = 3/4$ and $Y = - 1/2$  for the doublet sleptons,
and
$C_2 = 0$ and $Y = 1$ for the right handed selectrons.
$\cal S$ is given by,
\begin{eqnarray}
{\cal S} &=& \tr\left[ Y_{i} m_{i}^2 \right] =
m_{H_u}^2 - m_{H_d}^2
+ \tr\left[
m_{Q}^2
-m_{L}^2
-2m_{\bar{U}}^2
+m_{\bar{D}}^2
+m_{\bar{E}}^2
\right]\ .
\end{eqnarray}
The purely gauge mediated contributions to the above expression cancel at the messenger scale.
As we have discussed, the two-loop contributions to $m_{Q_3, \bar{T}}^2$
in Eq.\,(\ref{eq:Twoloop}) are  large and  positive for $y_t' \gtrsim 1$,
giving a negative ${\cal S}$.
Therefore, through the renormalization group equations,
the doublet sleptons become lighter at the low energy scale,
while the right-handed selectrons become heavier.%
\footnote{
The squark masses also receive the similar size of the renormalization group effects
from ${\cal S}$ with the signs depending on their $U(1)$ hypercharges.
}

\begin{figure}[t]
\begin{center}
  \includegraphics[width=.55\linewidth]{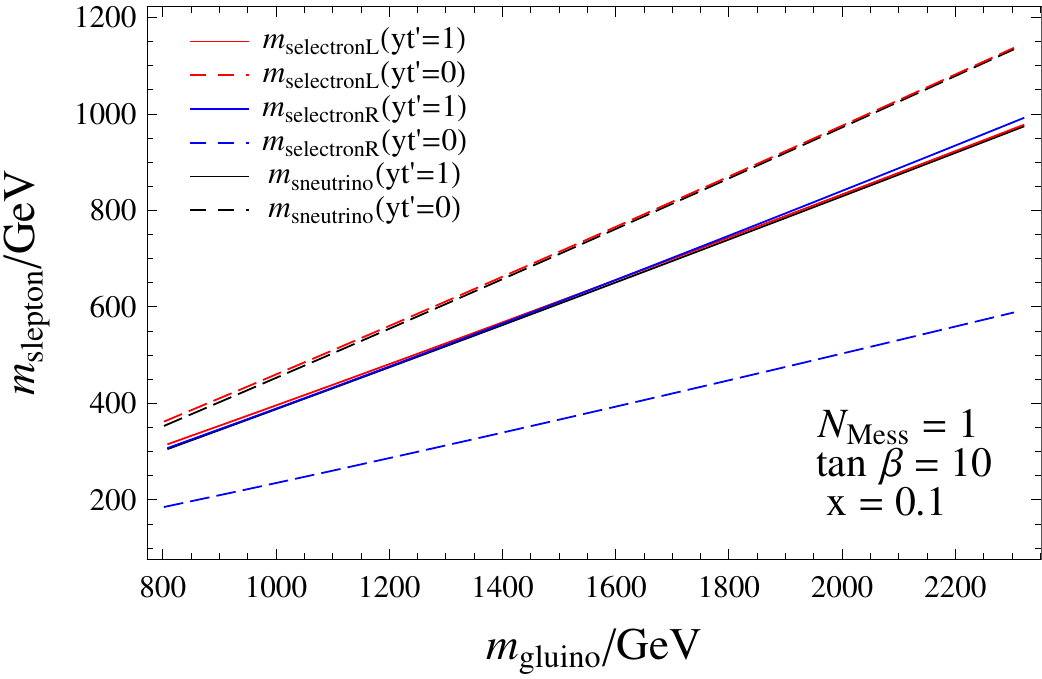}
\caption{\sl \small
The masses of the sleptons in Type-II models (solid lines) for $y_t' = 1$ and  $x = 0.1$
and for conventional models of gauge mediation (dashed lines).
The red and blue lines correspond to the masses of two sleptons.
The black lines correspond to the sneutrino mass.
The figure shows that the right-handed selectron mass is larger than that in
the conventional models, while the left-handed selectron
and the sneutrino masses are smaller than those in the conventional models.
The degeneracy of the slepton masses are just coincidence due to the choice
of the parameter.
In this figure, we have used $m_t = 173.2$\,GeV.
}
\label{fig:slepton}
\end{center}
\end{figure}

In Fig.\,\ref{fig:slepton}, we show the slepton masses for $y_t' = 1$
and $x  = 0.1$.
The figure shows that the right-handed selectrons are heavier than those in
conventional gauge mediation models, while the left-handed selectron
and sneutrino are lighter than expected.
Theis peculiar slepton mass spectrum also provides an important clue
for  probing Type-II gauge mediation models.

In the above analysis, we considered relatively large Yukawa coupling, $y_{t}'\simeq1$.
One might think that such a large Yukawa coupling has a Landau pole problem
well below the GUT scale.
A Landau pole in the RG running of the Yukawa coupling,
however, does not occur. First of all, the Yukawa coupling $y_t'$ is only important for renormalization group scales larger then messenger scale, $M$.
For such large energy scales,  Eq.\,(\ref{eq:superpotential1}) is a valid description of the model.
Thus, if the other coupling constants $g$ and $g' $ are rather small, $g\sim g' \ll 1$,
the only relevant Yukawa interaction in the high energy theory is
\begin{eqnarray}
W = \tilde y_t \tilde H_u Q_L\bar{T}_R\ ,
\end{eqnarray}
which is just the Yukawa coupling constant of the top quarks. This coupling is related to the low scale parameters as follows
\begin{eqnarray}
\tilde{y}_t = \sqrt{ y_t^2 + y_t'^2}\ .
\end{eqnarray}
Since $y_t$ at the messenger scale is around $y_t \sim 0.6-0.8$, $\tilde y_t $ does not significantly exceed $\tilde y_t = 1$
even for $y_t' \simeq 1$. Therefore, the coupling constant $\tilde y_t$ will be perturbative
up to the GUT scale.

Finally, let us comment on the effects of the other parameters.
In the figures, we have taken $\tan\beta = 10$.
The Higgs mass prediction is, however, almost independent of the choice of $\tan \beta $
as long as $\tan \beta \gtrsim 10$.%
\footnote{
The large $\tan\beta$ is favorable to explain the observed discrepancy 
of the observed muon anomalous magnetic moment from the SM prediction\,\cite{Bennett:2006fi}.
A detailed analysis is in preparation\,\cite{EIY}.
}
We also find that the predicted Higgs boson mass is almost
unchanged if we change the sign of the $\mu$-term,
since the $\mu$-term contribution to the stop mixing is suppressed by $1/\tan\beta$.

\section{Conclusions and Discussions}
In this paper, we discussed more generic models of gauge mediation
where the messengers are allowed to mix with the Higgs doubles via a ``charged" coupling constant.
  Although the messengers couple to the MSSM matter fields at the weak scale,
the flavor structure of these couplings is determined and is aligned to
the MSSM Yukawa couplings. Because of this alignment, the mixing
does not cause any serious flavor changing neutral current problems.

A distinguishing feature of these models with Higgs-messenger mixing is a one- and two-loop soft supersymmetry breaking mass for the sfermions proportional to the Yukawa couplings. Because of the hierarchy of the the Yukawa couplings, the most important contribution is to the stops. This important one-loop stop mass together with a relatively large stop A-term gives a unique superparticle and Higgs boson mass spectrum.
Particularly, we showed that the lightest Higgs boson can be as heavy as $125$\,GeV, for
example, with a gluino mass of around $2$\,TeV.
This is a remarkable difference from the situation
in conventional gauge mediation models where the lightest
Higgs boson mass cannot exceed $120$\,GeV for a gluino
mass in reach of the LHC.
Notice that the particle content in our model is the same as in the minimal gauge mediation.
We also found that in regions with an enhanced Higgs boson mass, the stops are heavier
than those predicted by conventional gauge mediation.

We should also note that the predictions of heavy
stops, sleptons with a large left-right mass splitting, and a heavy lightest Higgs boson
mass is unique to these models. The Next-to MSSM (NMSSM) can also predict a relatively heavy lightest Higgs boson
(see for example Ref.\,\cite{Morrissey:2008gm}), but when combined with gauge mediation has the vanilla mass spectrum of conventional gauge mediation.
This is because the stop and slepton masses are unaffected by the additional fields of the NMSSM.
Therefore, the interplay between the SUSY particle searches and the Higgs searches
are quite important for probing Type-II gauge mediation. We should note here that if we combine our mechanism with the NMSSM,
the Higgs mass may be raised up to 140 GeV for example \cite{EIY}.

Finally, let us comment on other interesting features of Type-II gauge mediation models.
In this paper, we were mainly concerned with the parameter space
which maximized the lightest Higgs boson mass, i.e. $m_{h^0}\gtrsim 120$\,GeV.
It should also be noted that the LEP bound on the Higgs boson mass,
$m_h^0 \gtrsim 114$\,GeV, is easily satisfied even for a gluino mass,
$m_{\rm gluino} \lesssim 1$\,TeV in Type-II gauge mediation (see Fig.\,\ref{fig:Higgs}).
Therefore,  in this case, SUSY particles may be discovered in the near future at the LHC experiments.

As another interesting possibility for these models,
the $\mu$-term can be relatively small
while keeping the Higgs boson mass above the current lower limit, $m_{h^0}\gtrsim 114$\,GeV.
This peculiar mass spectrum is made possible by the negative one-loop contributions
to the stops mass and the relatively large $A$-terms for $x\simeq 1$ 
(for related discussion see Refs.\,\cite{Kitano:2006gv,Asano:2010ut}).

\subsection*{Note added}
After the present paper was posted on arXiv (arXiv:1107.3006), ATLAS and CMS experiments reported 
excesses in the Higgs boson searches in the mass range of $m_{h^0} = 120-140$\,GeV
at a confidence levels close to 3$\sigma$\,\cite{LHC}.
Such a heavy lightest Higgs boson mass favors Type-II gauge mediation with $y_t' \simeq 1$
if the SUSY particles are also found at LHC experiments in the near future.
\section*{Acknowledgements}
We would like to thank Matt Sudano and Satoshi Shira for useful discussions.
This work was supported by the World Premier International Research Center Initiative
(WPI Initiative), MEXT, Japan.
The work of T.T.Y. was supported by JSPS Grand-in-Aid for Scientific Research (A)
(22244021).

\end{document}